# Data-Driven Security Assessment of the Electric Power System

Seyedali Meghdadi
Faculty of IT
Monash University
Melbourne, Australia
seyedali.meghdadi@monash.edu

Guido Tack
Faculty of IT
Monash University
Melbourne, Australia
guido.tack@monash.edu

Ariel Liebman
Faculty of IT
Monash University
Melbourne, Australia
ariel.liebman@monash.edu

*Abstract*— The transition to a new low emission energy future results in a changing mix of generation and load types due to significant growth in renewable energy penetration and reduction in system inertia due to the exit of ageing fossil fuel power plants. This increases technical challenges for electrical grid planning and operation. This study introduces a new decomposition approach to account for the system security for short term planning using conventional machine learning tools. The immediate value of this work is that it provides extendable and computationally efficient guidelines for using supervised learning tools to assess first swing transient stability status. To provide an unbiased evaluation of the final model fit on the training dataset, the proposed approach was examined on a previously unseen test set. It distinguished stable and unstable cases in the test set accurately, with only 0.57% error, and showed a high precision in predicting the time of instability, with 6.8% error and mean absolute error as small as 0.0145.

## I. INTRODUCTION

### A. Background and motivation

Static security limits are usually calculated off-line based on the worst-case scenarios. These limits are typically used throughout the year in the dispatch program; however, most of the time during the year, the system condition is not close to the worst-case scenario used to obtain the limits. Therefore, the dispatch results are not efficient at all the hours throughout the year [1]. Moreover, system security events, i.e. severe transient disturbances, involve large excursions of generator rotor angles, power flows, and other system variables, determined by non-linear characteristics of power system components. It forces the system to operate outside of the defined technical limits. As a result, a large set of security events are critically required to be taken into account to make sure that the system settles to new operating conditions such that no physical constraints are violated [2, 3]. This indicates the computational burden of a thorough transient stability assessment while planning the operation of the system. Transient stability assessment methods developed since the mid-1920s fall into the following categories:

**Time domain simulation** The time domain numerical integration method has been the universal method for transient stability analysis, where numerous algebraic and differential equations are iteratively solved in short time steps. However, it cannot address all control issues because it is computationally expensive.

**Direct or hybrid Methods** Direct methods aim to accelerate the assessment process by simplifying the complex model of a multi-machine power system. They include transfer energy function based methods, the (extended) equal area criterion (EEAC), and the single machine equivalent (SIME) paradigm [4-6]. Although the simplifications were justifiable when the methods were introduced, they are not very suitable for today's systems as they underestimate the security of the system and therefore lead to more expensive operation scenarios.

**Data-driven methods** Data-driven methods develop models to obtain mapping relations between inputs (system variables) and outputs (stability status). The core idea is that a comprehensive simulation of faults provides useful information regarding vulnerable locations in the system during different operation scenarios, and therefore reassures safe stability margins. ML approaches are widely accepted as an alternative way to address complex problems in modern power grid stability control and have been applied in various transient stability assessment (TSA) and control designs [7].

### B. Proposed approach

In this study, it is assumed that power electronic devices will not affect transient stability; hence, it is mainly determined by online synchronous generators in the grid. The algorithm proposed in this paper highlights weak lines, i.e. the lines with a high number of unstable instances, in a small group of adjacent synchronous generators as well as dangerous contingencies, i.e. contingencies with a very short time of instability, i.e. the time it takes for the system to become unstable after fault occurrence, for any given generation scenario. Since it decomposes the stability of a large system into a number of sub-systems (groups of adjacent synchronous generators), it drastically shrinks the size of the required dataset for the training of the classifiers. Moreover, it is flexible to changes in the topology of the grid since each change (e.g. retirement of a power plant) will only affect one trained classifier; hence, it is computationally efficient. On the other hand, by introducing the concept of time of instability it not only accurately avoids absolute dangerous scenarios, i.e. operational scenarios where the system is at its lowest security level with a short time of instability, but also it will reduce the operation cost of power system as it enables choosing scenarios with fewer number of online fossil fuel power plants by utilising the contribution of energy storage systems (ESS) to system stability during the frequency response sequence (e.g. inertia emulation). For instance, assume τ is the time constant representing the time required for detection of a contingency and activation of an ESS. The algorithm can allow the operation of scenarios with a larger time of instability, such as multi-swing instability, where the time of instability is greater than τ.

### C. Literature review

In conventional stability analysis, only a small number of worst-case critical conditions are analysed, using time domain simulations, and then the results are generalized to all possible credible operating conditions [8]. Moreover, the selection of critical contingencies is mostly based on historical performance and the expert's experience [9-11].

Influential methods such as SIME and EEAC were introduced to address stability assessment more efficiently [12]. However, with the advent of advanced information technology, new methods with faster response time and higher accuracy were developed. As an alternative to direct methods, data-driven approaches such as machine learning-based frameworks, e.g. neural networks (NN) [13], support vector machines (SVM) [14], or decision trees (DT) [15] were exploited to detect stability status. The highly nonlinear, computationally expensive, and time-constrained nature of transient stability analysis makes ML approaches well suited for this application. It is important to notice that the ML techniques should only use the static variables available from ACOPF. It would not be practical if the algorithm required pre-fault dynamic (state variables) or post-fault variables as inputs since they impose huge pre-processing calculations [16-19].

Some other researchers used deep learning (DL) approaches exploiting its effective automatic feature selection capability. For instance, authors in [20] use Extreme Learning Machine (ELM), a DL based framework where large datasets are required, to increase the accuracy of classification. ELM has also been used for fast stability scanning, amongst other methods [21]. The authors in [22] suggest pattern discovery (PD) for knowledge extraction since these classifiers tend to be sensitive to little changes in the data. It exploits a distance-based feature estimation algorithm called RELIEF to statically identify critical generator features. The credibility of the algorithms is not reported sufficiently high, as they introduce a new class as a response to inputs, especially where little data is available, given the credibility of the results varying between 89.25% and 94.48% depending on the fault location. Therefore, the number of calculations and complexity of the proposed algorithms sound unnecessary, resembling a black box.

Concisely, the main deficiency of the applied deep learning algorithms is that they consider no prior understanding of power system operation and therefore rely solely on large-sized collected data and use of complex algorithms.

*D. Contributions*

The contributions made in this paper are two-fold:

*a) Critical line method*

A novel method, called *critical line* is proposed in this paper, enabling the use of straightforward machine learning techniques. It will avoid the use of complex mathematical calculations finding critical machines in other algorithms by breaking the colossal stability assessment problem into a number of smaller sub-problems. It is based on the fact that the disturbance effect will spread from the trip location to the rest of the network according to the network impedances and generator inertias [23]; therefore, the interaction of the dynamics of synchronous generators are more effective within adjacent generators. Consequently, in this method, all synchronous generators in the system will be divided into small groups depending on their electrical distance and inertia. Then, the stability of each group will be studied separately. Within each group, critical lines (weak lines), i.e. lines with a large number of unstable instances for a comprehensive set of generation scenarios based on the average power transfer and the topology of the system, will characterize the response of the group of generators to disturbances. Therefore, it can clearly describe how the instability manifests in each group of generators. On the other hand, inter-area oscillations amongst defined groups have a time scale in the range of tens of seconds [3]. Therefore, they have not been considered in this study

We identified the following characteristics within unstable instances and propose the solution accordingly:

- The relation between line loading and the number of instabilities, i.e. the number of generation scenarios that make the system unstable after the occurrence of a fault on the line, on each line for any given generation scenarios is very significant.

- Instabilities in lines with fewer unstable instances (stiff lines) are subclasses of the line with the highest number of unstable instances (weak lines), meaning the same generation scenario also makes the system unstable after fault occurrence at the weakest line.

- Time of instability is greater in stiff lines compared to weak lines, meaning those instances are less dangerous at those locations.

Although this method could be applied to all the lines in the system, the data gathered from stiff lines would not be statistically significant. Therefore, we may assume once the weak lines are detected, assessing the stability at stiff lines do not add notable information about the behaviour of the sub-systems. Moreover, it improves the scalability of the algorithm when applied to larger systems as the supervised learning process is only executed for the critical lines at each small group of generators. Therefore, it facilitates the use of machine learning algorithms to enable faster and more reliable decision-making during planning.

*b) Two-stage classification algorithm*

The algorithm we propose exploits two-level classification for all of the critical lines in the system to predict the stability index as well as the time of instability. This machine learning tool will be added to the optimal unit commitment problem (aka. short-term planning). The ML tool will provide stability status of any calculated feasible operation scenario, replacing the computationally expensive time domain simulations. It is composed of a binary and a non-binary classifier to detect stability index, i.e. distinguish stable cases from unstable ones with a very high level of precision, and the time of instability respectively.

Transient stability in the presence of all of the system regulators is the outcome of numerous iterative numerical integration calculations. Therefore, mapping the measured pre-fault steady-state parameters to stability status and time of instability at the same time is difficult while the training dataset is imbalanced. Recently, different approaches to feature selection have been introduced to reduce the number of input features [24]. The proposed approach uses the following parameters as inputs to the machine learning tool: real and reactive power of each load, real and reactive power of each generator, voltage angle and voltage magnitude at the ends of each line, and final objective function value indicating the minimum cost of power generation for the load scenario.

## II. INTELLIGENT TRANSIENT STABILITY ASSESSMENT

An intelligent system can provide a high level of ability to uncover salient but previously unknown characteristics of a system. It is used to map operation conditions to the security status when it is properly trained. Fig. 1 depicts the overall process to prepare a transient stability assessment (TSA) tool for each critical line in the system. The core of a TSA is the classifier which immediately distinguishes the transient stability of a power system, once an operating condition is fed. A key feature of a classifier is its generalization ability, referring to its ability to give reliable and accurate predictions using previously unseen operating conditions. In supervised learning, a training set consists of a group of operating conditions obtained from ACOPF, the corresponding stability labels obtained using time domain simulations.

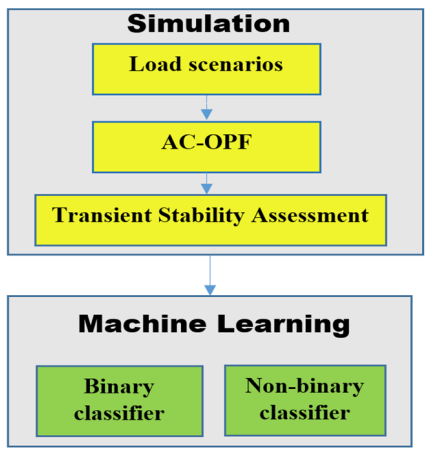

Fig. 1- Block diagram of the proposed machine learning solution

The main steps to generate the training and validation datasets are as follows:

### A. Simulation

**Load scenarios** Postulated load scenarios are produced around the benchmark load values of the test case. Furthermore, they not only should reflect the possible operating region, but also should be able to push the system to its stability margins. Consequently, there will be enough unstable instances, enabling generators and transmission lines to operate close to their thermal limits in the presence of the power system controllers.

**AC Optimal Power Flow** AC optimal power flow (ACOPF) is used to create generation values for postulated load scenarios. It generally aims at minimizing the cost function while ensuring the operation of the system within its allowed limits. MATPOWER is used in this study to solve the AC-OPF problem[25].

**Transient stability assessment** A detailed simulation including all system components is created and a training dataset is prepared to be used for supervised learning in the next step. Real and reactive power of loads and generators as well as the angle and magnitude of bus voltages, extracted from the ACOPF study, are supplied to the Matlab/ Simulink root-mean-square (RMS) simulation model.

### B. Machine learning

A two-stage classifier is trained to predict the behaviour of small sub-systems, i.e. each group of generators, with respect to a contingency at critical lines. It will predict both the stability status and the time of instability of the sub-systems. Each classifier predicts the dynamic behaviour regarding a contingency happening within the sub-system. Then, all classifiers will be put in parallel to depict the big picture by representing the behaviour of the large system.

## III. APPLICATION TO AN IEEE TEST CASE

In this section, the proposed intelligent system is created, tested, and then results are reported. The Western Electric Coordinating Council (WECC) nine-bus benchmark is used to conduct the experiments (Fig. 2) [26]. This system, while small, is large enough to be nontrivial and thus permits the illustration of a number of stability concepts and results. We will consider this system as one group of generators.

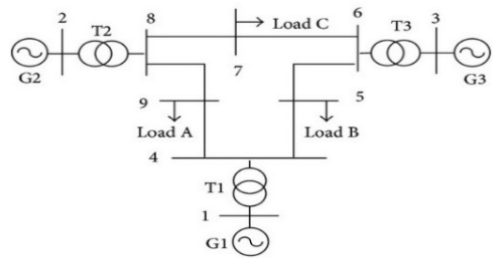

Fig. 2- One-line diagram of WECC system

To provide an unbiased evaluation of the trained model, two different data sets are required to train and test the classifiers. Supervised learning approach will be conducted on the training dataset, and the performance of the trained algorithm will be examined on a previously unseen test set. For the training dataset, loads can randomly take a coefficient within 0.3 and 1.7 of the benchmark values subject to the following constraint: the summation of all coefficients must be within a bandwidth ensuring both high and low loading scenarios. Fig. 3 shows the scenarios produced for training purposes (455 scenarios). The x-axis shows the generation scenario sequence number and y-axis depicts load value in MW.

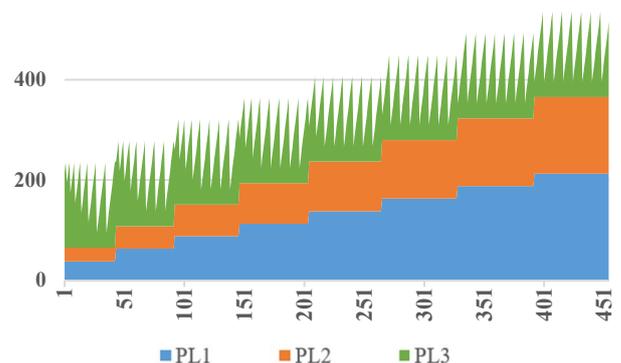

Fig. 3- Proposed load variation for the training dataset

Each load in the validation data set can take a coefficient from 0.25 to 1.85, where the summation of all coefficients must be within the same limit as in the training dataset. The performance of the classifiers is guaranteed while they interpolate. However, comparing load coefficients of training and validation dataset indicates validation dataset is designed to cautiously test inter and extrapolation capability of the trained classifiers. Fig. 4 shows 699 produced validation scenarios.

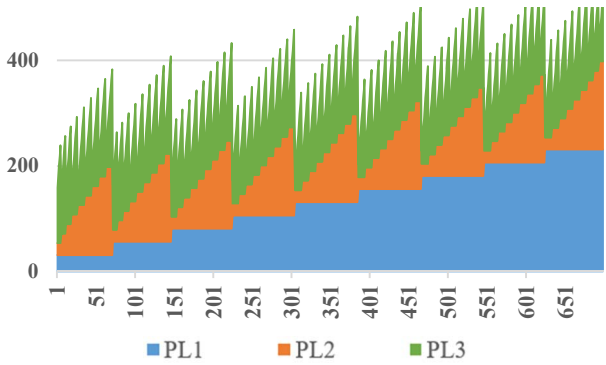
Fig. 4- Postulated load variation for validation dataset

Then, for each scenario, real and reactive power of each load along with all generators' real and reactive power, voltage magnitude and voltage angle of both ends of the line, and the cost of the generated power from the ACOPF algorithm, are extracted. Fig. 5 indicates ACOPF solution values regarding active power output of each generator for postulated load variations of the training dataset. It was observed that the number of instabilities for the most critical line of the test system, line 5-7, are 104 out of 455 scenarios (23%) for the training dataset and 198 out of 699 scenarios (28%) for the validation dataset.

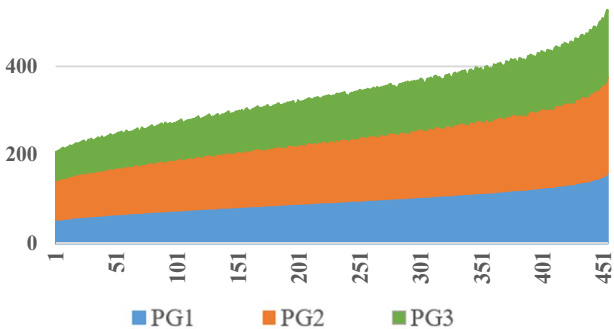
Fig. 5- Generation values of ACOPF

The training is conducted with the help of Matlab Neural Pattern Recognition and Classification Learner toolboxes. Moreover, the fine tuning of the hyperparameters are performed using Bayesian Optimisation method to iteratively develop a global statistical model of the unknown objective function, achieving maximal effectiveness in a fixed time [27]

*A. Binary classifier*

A neural network with one hidden layer of 10 hidden neurons is trained using a scaled conjugate gradient backpropagation algorithm to capture the high nonlinear relations between the inputs and binary output, indicating a negative class (unstable scenario) or positive class (stable scenario) respectively. If the output of the NN falls between zero and one, the stability should be determined based on the confidence of the NN output. The distance of the values between zero and one to each end of the range can be used as a measure of confidence of the neural networks performance and will be used during further development of the algorithm. Fig. 6, indicating the confusion matrices of the proposed binary classifier, proves a high accuracy level, where each cell value is shown in percentage. It is important to highlight that due to the high performance of the trained neural network, there was no need to use deep network structures with multiple hidden layers.

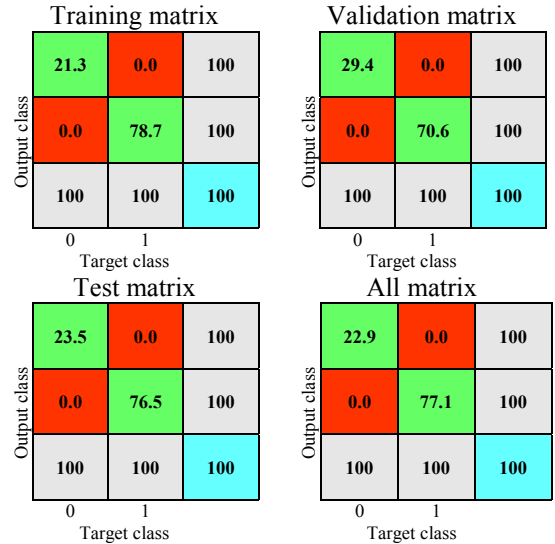
Fig. 6- confusion matrices of trained neural networks binary classifier

*B. Non-binary classifier*

The multiclass classifier needs to predict the time of instability with a very low level of error, even though it can only be trained on a small number of unstable instances. Therefore, estimating the time of instability is a difficult task. The data are standardized and the Kernel scale is chosen using heuristic procedure Then, the predictive accuracy of various fitted model was tested. Afterwards, fine tuning of the most accurate algorithms was performed to improve their efficiencies leading to find the best fit. It is important to acknowledge that these tunings are conducted both manually, such as cost of misclassification, and automatically using Bayesian optimisation, such as kernel scale and box constraint value. The accuracy of some of the tested classifiers is reported in Table 1. In the end, a quadratic SVM classifier with second order polynomial function, and 20-fold cross-validation proved to be the most accurate multi-class classifier with the highest accuracy level.

Table 1- The accuracy of multi-class classifiers

| Classifier | Category | Accuracy (%) |
| --- | --- | --- |
| Decision tree | medium | 58.7 |
| Decision tree | coarse | 61.5 |
| Discriminant | Linear | 73.1 |
| SVM | Quadratic | 76 |
| SVM | Medium Gaussian | 62.5 |
| SVM | Coarse Gaussian | 53.8 |
| KNN | Fine | 54.8 |
| KNN | Cosine | 56.7 |
| KNN | Cubic | 50 |
| Ensemble | Bagged trees | 54.8 |
| Ensemble | Boosted trees | 57.7 |

Fig. 7 depicts the confusion matrix of a quadratic SVM, showing an acceptable accuracy. It can be seen that most miss-classifications were due to the trade-off between the size of the collected data set and accuracy. For instance, a true class of 2 is miss-classified as 1.7 since only 2 data points where provided.

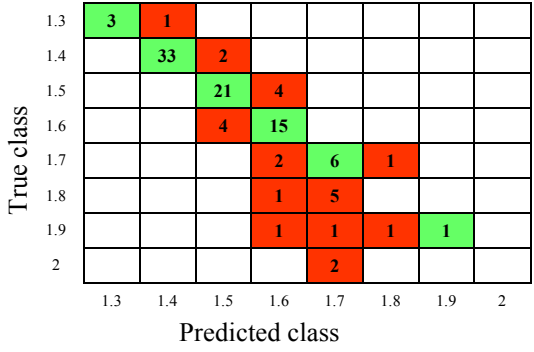

Fig. 7- Quadratic SVM non-binary classifier confusion matrix

After the two classifiers are trained, postulated load variation for the validation dataset is used and the same process is conducted to generate the dataset. Then, the output of the proposed algorithm using the trained classifiers is compared to the actual response of the system, obtained from time domain simulations.

## IV. RESULTS

Regarding machine learning implementation and accuracy, for the 455 instances of training data set, the algorithms achieve 100% accuracy for binary classification (NN) and 76% accuracy for the non-binary classification (SVM) on training data set. To test the performance of the proposed algorithm, a validation dataset with 699 scenarios (different from training) was fed into the two-stage classification algorithm. It was observed that the binary classifier could distinguish stable and unstable cases accurately, with only 0.57% error. The non-binary classifier also performed a remarkable prediction in detecting the time of instability, where the predicted time of instability had 6.8% labelling error. It provided 100% credibility, e.g. it did not introduce a new class as a response to inputs, which is one of the drawbacks of deep learning methods especially where little data is provided. To better understand the performance of the non-binary classifier, the mean absolute error (MAE) is calculated using equation (1).

$$MAE = \frac{1}{N}\sum_{i=1}^{N}|y_i^{predicted} - y_i^{actual}| \quad (1)$$

The MAE of the proposed approach is as small as 0.0145, showing the accurate performance of the non-binary classifier. Moreover, the mean value of actual versus predicted the time of instability are 1.61 and 1.57 respectively. Also, the variance of actual versus predicted the time of instability within misclassified are 0.03 and 0.02. These incidents are very similar. Fig. 8 shows a scatter plot indicating the error of prediction ( Error = $|y_i^{predicted} - y_i^{actual}|$) for all 48 misclassified incidents. A small difference between the predicted time of instability (in orange) and the actual time of instability (in blue) is evident, where the error value is plotted in green. There is a trade-off between the accuracy of the predicted time of instability and the size of the gathered dataset. For instance, introducing previously learnt mislabelling will increase the accuracy; however, the aim of the authors was to keep the size of the required dataset as small as possible and to minimise the manual interferences while still maintaining justifiable accuracy.

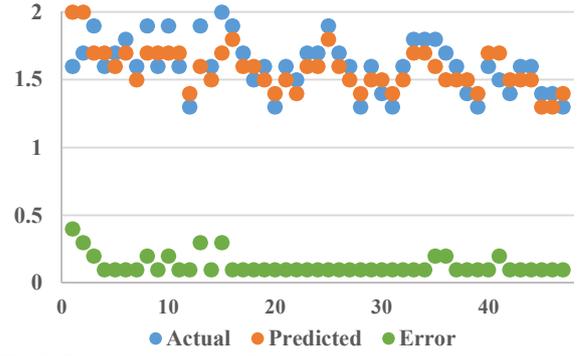

Fig. 8- Scatter plot of time of instability

## V. CONCLUSION

The highly nonlinear, computationally expensive, and time constrained nature of the transient stability control problem makes machine learning techniques well-suited for this application. This paper addressed the bottleneck of using AI based approaches in the presence of control devices – a step forward compared to previous research with uses simplified versions of the system where there is no control unit available. This new approach will help not to underestimate the security of the system and can help decrease electricity costs for costumers, as it increases the available search space for the optimal power flow problem.